\documentclass[fleqn,twoside]{article}
\usepackage{espcrc2}

\usepackage{graphicx}

\newcommand{\lsim}{\mathrel{\rlap{\lower4pt\hbox{\hskip0pt$\sim$}}
\raise1pt\hbox{$<$}}}           
\newcommand{\gsim}{\mathrel{\rlap{\lower4pt\hbox{\hskip0pt$\sim$}}
\raise1pt\hbox{$>$}}}           

\newcommand{\sfrac}[2]{\mbox{\footnotesize $\frac{#1}{#2}$}}

\newcommand{\AmS}{{\protect\the\textfont2
  A\kern-.1667em\lower.5ex\hbox{M}\kern-.125emS}}

\title{Goldstone Boson's Valence-Quark Distribution\thanks{I thank M.B.~Hecht and
S.M.~Schmidt for useful communications.  This work was supported by: the US
Department of Energy, Nuclear Physics Division, under contract
no.~\mbox{W-31-109-ENG-38}; and in part by the Deutsche Forschungsgemeinschaft
through the Mercator Visiting Professor Programme under contract no.
Ro~1146/3-1.}}

\author{C.D.~Roberts\\[1ex]
Physics Division, Argonne National Laboratory, Argonne IL 60439, USA;\\
and Fachbereich Physik, Universit\"at Rostock, D-18051 Rostock, Germany}

\begin{document}

\begin{abstract}
Dynamical chiral symmetry breaking (DCSB) is one of the keystones of low-energy
hadronic phenomena. Dyson-Schwinger equations provide a model-independent
quark-level understanding and correlate that with the behaviour of the pion's
Bethe-Salpeter amplitude.  This amplitude is a core element in the calculation
of pion observables and combined with the dressed-quark Schwinger function
required by DCSB it yields a valence-quark distribution function for the pion
that behaves as $(1-x)^2$ for $x\sim 1$, in accordance with perturbative
analyses. This behaviour can be verified at contemporary experimental
facilities.
%
%
\vspace{1pc}
\end{abstract}

\maketitle

\section{DCSB and the Gap Equation}
Dynamical chiral symmetry breaking (DCSB) is a signature characteristic of the
strong interaction spectrum and it is a feature of QCD that can only be
understood via nonperturbative analysis.\footnotemark[1]
%
In this connection one particularly insightful tool is the QCD gap equation;
i.e., the Dyson-Schwinger equation (DSE)~\cite{cdragw} for the dressed-quark
self-energy:
\begin{eqnarray}
\nonumber
\lefteqn{S(p)^{-1}  =  Z_2 \,(i\gamma\cdot p + m_{\rm bare})}\\
&&  +\, Z_1 \int^\Lambda_q \, g^2 D_{\mu\nu}(p-q) \frac{\lambda^a}{2}\gamma_\mu
S(q) \Gamma^a_\nu(q,p) \,. \label{gendse}
\end{eqnarray}
In Eq.~(\ref{gendse}): $D_{\mu\nu}(k)$ is the renormalised dressed-gluon
propagator; $\Gamma^a_\nu(q;p)$ is the renormalised dressed-quark-gluon vertex;
$m_{\rm bare}$ is the $\Lambda$-dependent current-quark bare mass that appears
in the Lagrangian; and $\int^\Lambda_q := \int^\Lambda d^4 q/(2\pi)^4$
represents mnemonically a {\em translationally-invariant} regularisation of the
integral, with $\Lambda$ the regularisation mass-scale.  Also,
$Z_1(\zeta^2,\Lambda^2)$ and $Z_2(\zeta^2,\Lambda^2)$ are the
quark-gluon-vertex and quark wave function renormalisation constants, which
depend on the renormalisation point, $\zeta$, and the regularisation
mass-scale, as does the mass renormalisation constant:
\begin{equation}
\label{Zmass} Z_m(\zeta^2,\Lambda^2) =
Z_4(\zeta^2,\Lambda^2)/Z_2(\zeta^2,\Lambda^2) ,
\end{equation}
with the renormalised mass given by $ m(\zeta) := m_{\rm
bare}(\Lambda)/Z_m(\zeta^2,\Lambda^2)$.

The solution of Eq.~(\ref{gendse}) has the form
\begin{eqnarray}
\label{sinvp} S(p)^{-1} & = & i \gamma\cdot p \,A(p^2,\zeta^2) +
B(p^2,\zeta^2)\\
       & = & \frac{1}{Z(p^2,\zeta^2)}\left[ i\gamma\cdot p +
        M(p^2,\zeta^2)\right]\,,
\end{eqnarray}
where the functions $A(p^2,\zeta^2)$, $B(p^2,\zeta^2)$ express the effects of
dressing induced by the quark's interaction with its own gluon field.
Equation~(\ref{gendse}) must be solved subject to a renormalisation condition,
and in QCD it is practical to impose the requirement that at a large spacelike
$\zeta^2$
\begin{equation}
\label{renormS} \left.S(p)^{-1}\right|_{p^2=\zeta^2} = i\gamma\cdot p +
m(\zeta)\,.
\end{equation}
\footnotetext{These observations are also true of confinement but
model-independent statements harder to make.}

A weak coupling expansion of the DSEs reproduces every diagram in perturbation
theory and in connection with Eq.~(\ref{gendse}) this means that at large
spacelike-$p^2$ the solution for massive quarks is, in Landau gauge,
\begin{equation}
\label{masanom} M(p^2) = \frac{\hat m} {\left(\rule{0mm}{1.2\baselineskip}
\sfrac{1}{2}\ln\left[p^2/\Lambda_{\rm QCD}^2 \right]\right)^{\gamma_m}}\,,
\end{equation}
$\gamma_m= 12/(33-2 N_f)$, where $\hat m$ is the
renormalisation-point-independent current-quark mass, and the mass
renormalisation constant, Eq.~(\ref{Zmass}), is $Z_m(\zeta^2,\Lambda^2) =
\left[ \alpha(\Lambda^2)/\alpha(\zeta^2)\right]^{\gamma_m}$.  At one-loop
$Z_2\equiv 1 \equiv Z(p^2)$.  In perturbation theory each contribution to the
dressed-quark mass function is proportional to the current-quark mass and hence
$M(p^2)\equiv 0$ in the chiral limit, which is unambiguously defined by
\mbox{$\hat m = 0$}~\cite{mrt98}.

Dynamical chiral symmetry breaking is the appearance of a
$B(p^2,\zeta^2)\not\equiv 0$ solution of Eq.~(\ref{gendse}) in the chiral
limit.  This guarantees a nonzero value of the vacuum quark
condensate~\cite{mrt98}:
\begin{equation}
\label{qbq0} \,-\,\langle \bar q q \rangle_\zeta^0 = Z_4(\zeta^2,\Lambda^2)\,
N_c {\rm tr}_{\rm D}\int^\Lambda_q\,
        S^{0}(q,\zeta)\,,
\end{equation}
where ${\rm tr}_D$ identifies a trace over Dirac indices only and the
superscript ``$0$'' indicates the quantity was calculated in the chiral limit.
(NB.\ The factor of $Z_4$ here guarantees that the condensate is
gauge-parameter- and cutoff-independent.  Its omission yields a formula that is
incorrect.)

As I have just observed, $B(p^2,\zeta^2)\not\equiv 0$ is impossible in
perturbation theory.  Hence a nonperturbative analysis of QCD's gap equation is
required to explore this possibility.  To arrive at model-independent
conclusions a systematic, symmetry-preserving truncation scheme for the
$n$-point functions in the gap equation must be used.

\subsection{Goldstone's Theorem}
At least one such scheme is known~\cite{truncscheme} and it has been used to
good effect.  For example, the pion's Bethe-Salpeter amplitude has the general
form
\begin{eqnarray}
\nonumber
\lefteqn{ \Gamma_\pi^j(k;P) = \tau^j \gamma_5 \left[ i E_\pi(k;P)
+ \gamma\cdot P F_\pi(k;P) +\right.}\\
\nonumber & &   \left. +\,  \gamma\cdot k \,k \cdot P\, G_\pi(k;P) +
\sigma_{\mu\nu}
k_\mu P_\nu \, H_\pi(k;P) \right],\\
&& \label{genpibsa}
\end{eqnarray}
where $P$ is the pion's total momentum and $k$ is the relative momentum between
the bound state's constituents.  The scheme of Ref.~\cite{truncscheme}
guarantees that the axial-vector Ward-Takahashi identity (here written for
$\hat m = 0$, $k_\pm= k\pm P/2$):
\begin{eqnarray}
\nonumber\lefteqn{-i P_\mu \Gamma_{5\mu}^j(k;P)  }\\
& &  = S^{-1}(k_+)\gamma_5\frac{\tau^j}{2} +  \gamma_5\frac{\tau^j}{2}
S^{-1}(k_-) \,, \label{avwtich}
\end{eqnarray}
where
\begin{eqnarray}
\nonumber \lefteqn{\Gamma_{5 \mu}^j(k;P)  = \frac{\tau^j}{2} \gamma_5 \left[
\gamma_\mu F_R(k;P) + \right.}\\
&&   \nonumber \left. \gamma\cdot k k_\mu G_R(k;P) - \sigma_{\mu\nu} k_\nu
H_R(k;P) \right] + [\ldots] ,\\
\label{truavv}
\end{eqnarray}
with $F_R(k;P)$, $G_R(k;P)$, $H_R(k;P)$ regular at $P^2=0$, is satisfied at
every order.  This identity, which is fundamental to the successful application
of chiral perturbation theory, leads to the following quark-level
Goldberger-Treiman relations~\cite{mrt98}, which are exact in QCD:
\begin{eqnarray}
\label{bwti} f_\pi^0 E_\pi(k;0)  &= &  B(k^2)\,, \\
\label{fwti}
 F_R(k;0) +  2 f_\pi^0 F_\pi(k;0)                 & = & A(k^2)\,, \\
\label{gwti}  G_R(k;0) +  2 f_\pi^0 G_\pi(k;0)    & = & 2 A^\prime(k^2)\,, \\
H_R(k;0) +2 f_\pi^0 G_\pi(k;0) & = & 0\,, \label{hwti}
\end{eqnarray}
where $f_\pi^0$ is the pion's chiral-limit leptonic decay constant, obtained in
general from
\begin{eqnarray}
\nonumber \lefteqn{\delta^{ij} f_\pi P_\mu = }\\
&& Z_2\int^\Lambda_q \! {\rm tr}\left[\frac{\tau^i}{2} \gamma_5 \gamma_\mu
S(q_+) \Gamma_\pi^j(q;P) S(q_-)\right], \label{caint}
\end{eqnarray}
with the factor of $Z_2$ on the right-hand-side guaranteeing that $f_\pi$ is
independent of the renormalisation point, cutoff and gauge parameter. (A
formula without $Z_2$ is incorrect.)  Equations~(\ref{bwti})--(\ref{hwti}) form
an essential part of the proof~\cite{mrt98} that if, and only if, a theory
exhibits DCSB then the homogeneous, isovector, pseudoscalar Bethe-Salpeter
equation has a massless, $P^2=0$, solution and the axial-vector vertex is
dominated by the pion pole for $P^2\simeq 0$; i.e., a proof of Goldstone's
theorem in QCD.

\subsection{A Pseudoscalar Meson Mass Formula}
Similar analysis of the $\hat m \neq 0$ axial-vector Ward-Takahashi identity
yields the following mass formula for flavour nonsinglet pseudoscalar mesons:
\begin{equation}
\label{massform}
f_H^2 \, m_H^2 = - \langle \bar q q \rangle^H_\zeta \, {\cal M}^H_\zeta \,,
\end{equation}
where: ${\cal M}^H_\zeta = m^\zeta_{q_1} + m^\zeta_{q_2}$ is the sum of the
current-quark masses of the meson's constituents; $f_H$ is the meson's leptonic
decay constant, obtained via obvious analogy with Eq.~(\ref{caint}); and the
in-meson condensate is defined via
\begin{eqnarray}
\nonumber\lefteqn{ i \frac{1}{f_H}\,\langle \bar q q \rangle^H_\zeta  = }\\
& &  Z_4
\nonumber \int_q^\Lambda \!
{\rm tr}\left[\left( \mbox{\small $\frac{1}{2}$} T^H\right)^{\rm t} \gamma_5
{\cal S}(q_+) \Gamma^H(q;P) {\cal S}(q_-)\right],\\
&&
\end{eqnarray}
with, e.g.,  $T^{\pi^+}= (\lambda^1 + i \lambda^2)/2$, where
$\{\lambda^j,j=1,\ldots, 8\}$ are the Gell-Mann matrices, and $(\cdot)^{\rm t}$
denoting matrix transpose.

A number of observations are now in order.  Equation~(\ref{bwti}) makes plain
that in the presence of DCSB the magnitude of $f_\pi$ is set by the
constituent-quark mass: $M(p^2\simeq 0)$.  The same is true of $\langle \bar q
q \rangle^H_\zeta$, which, in the chiral limit, is identical to the vacuum
quark condensate, Eq.~(\ref{qbq0}).  This hints at a corollary of
Eq.~(\ref{massform}): in the chiral limit it yields the
Gell-Mann--Oakes--Renner relation, a primary element in the application of
chiral perturbation theory to light-quark mesons.  Furthermore, while it is not
readily apparent from the material presented heretofore, the DSE derivation of
Eq.~(\ref{massform}) assumes nothing about the magnitude of
$m^\zeta_{q_1,q_2}$. Hence Eq.~(\ref{massform}) is a single mass formula that
unifies the light- and heavy-quark domains and it has consequently been used to
prove that the mass of an heavy pseudoscalar meson rises linearly with the mass
of its heaviest constituent~\cite{pieterheavy,mishaheavy}.

\section{Realising DCSB}
The truncation scheme of Ref.~\cite{truncscheme} also provides the foundation
for a phenomenological model that has been used to very good effect in
describing light-quark mesons~\cite{mr97,pieterrho,pieterpion,pieterother}. The
model is founded on the assumption that the leading order term in the
truncation of QCD's gap equation is well represented by a renormalisation group
improved rainbow approximation~\cite{mr97}:
\begin{eqnarray}
\nonumber\lefteqn{ S(p)^{-1} = Z_2 \,(i\gamma\cdot p + m_{\rm bare}) }\\
\nonumber  & +& \!\!\!\!\!\!
 \int^\Lambda_q \, {\cal G}((p-q)^2)\, D_{\mu\nu}^{\rm
free}(p-q) \frac{\lambda^a}{2}\gamma_\mu S(q) \frac{\lambda^a}{2}\gamma_\nu
\,, \\
&&  \label{ouransatz}
\end{eqnarray}
where $D_{\mu\nu}^{\rm free}(k)$ is the free gauge boson propagator.  In
Eq.~(\ref{ouransatz}), ${\cal G}(k^2)$ is an ``effective coupling:'' ${\cal
G}(k^2)=4\pi \alpha(k^2)$ for $k^2\gsim 1\,$GeV$^2$, where $\alpha(k^2)$ is the
strong running coupling constant, but the behaviour of ${\cal G}(k^2)$ for
$k^2< 1\,$GeV$^2$; i.e., at infrared length-scales ($\gsim 0.2\,$fm), is
currently unknown.

It is immediately obvious from Eq.~(\ref{gendse}) that the effective coupling
is a mnemonic for the contracted product of the dressed-gluon propagator and
dressed-quark-gluon vertex. Hence the infrared strength of the kernel in the
gap equation can be characterised by an interaction-tension:
\begin{eqnarray}
\nonumber
\lefteqn{ \sigma^{\Delta}:= }\\
&& \!\!\!\!\! \frac{1}{4\pi}\int_{\Lambda_{\rm QCD}^2}^{\Lambda_{\rm
pQCD}^2}\,dk^2\,k^2\,\left[\Delta(k^2)-\Delta(\Lambda_{\rm pQCD}^2)\right]
\label{IT}
\end{eqnarray}
with $\Delta(k^2)={\cal G}({k^2})/k^2$, and $\Lambda_{\rm pQCD} =
10\,\Lambda_{\rm QCD}$ the boundary above which perturbation theory is
unquestionably valid. Studies show~\cite{fredIR} that DCSB cannot occur for
$\sigma^{\Delta}\lsim 0.5\,$GeV$^2\sim 9\Lambda_{\rm QCD}^2$, which means that
the existence of a dynamically generated, $B\not\equiv 0$ solution of
Eq.~(\ref{gendse}) is impossible without a significant infrared enhancement of
the effective interaction. Reproducing observable phenomena requires
\begin{equation}
\label{sigmaphysical}
\sigma^{\Delta}\gsim 4\,{\rm GeV}^2\sim 70\,\Lambda_{\rm QCD}^2\,;
\end{equation}
i.e., a ten-fold enhancement over the critical value. The origin of this
enhancement, whether in the gluon or ghost vacuum polarisation, or elsewhere,
is currently unknown but is actively being sought (see, e.g.,
Refs.~\cite{fredIR,continuumgluon,latticegluon}).  That is a key to
understanding the source of confinement as opposed to simply representing it.

References~\cite{mr97,pieterrho,pieterpion,pieterother} employ a one-parameter
model for ${\cal G}(k^2)$, with that parameter, and the $u = d$- and
$s$-current-quark masses, tuned to fit the experimental values of $m_\pi$,
$f_\pi$, $m_K$.  This fit yields: the value of $\sigma^{\Delta} \approx
70\,\Lambda_{\rm QCD}^2$ in Eq.~(\ref{sigmaphysical});
\begin{equation}
\label{massuq} m_{u}^{1\,{\rm GeV}} = 5.5\,{\rm MeV}\,,\;
m_{s}^{1\,{\rm GeV}} = 124\,{\rm MeV}\,;
\end{equation}
and a value of the vacuum quark condensate:
\begin{equation}
- \langle \bar q q \rangle_{1\,{\rm GeV}}^0 = (0.242\,{\rm GeV})^3\,,
\end{equation}
for which a comparison with the calculated value
\begin{equation}
- \langle \bar q q \rangle_{1\,{\rm GeV}}^\pi = (0.245\,{\rm GeV})^3
\end{equation}
emphasises just how close one is to the chiral limit when dealing with pion
observables. In addition, the essential character of the dressed-quark
propagator predicted by the model, which underlies these condensate values, has
recently been confirmed in numerical simulations of
lattice-QCD~\cite{pmqciv,peteradelaide}.

Fixing the single parameter at this fitted value, many observables are
predicted and the model achieves a r.m.s.\ error over predicted quantities of
$\lsim 4\,$\%~\cite{revbasti}.  Furthermore, it is the only model to
predict~\cite{pieterpion} a behaviour for the pion's electromagnetic form
factor that agrees with the results of a recent Hall~C
experiment~\cite{volmer}.  The large-$Q^2$ behaviour of the form factor can be
obtained algebraically and one finds~\cite{mrpion} $Q^2 F_\pi(Q^2)=\,$const.,
up to logarithmic corrections, in agreement with the perturbative-QCD
expectation. This result relies on the presence of pseudovector components in
the pion's Bethe-Salpeter amplitude, which is guaranteed by the quark-level
Goldberger-Treiman relations in Eqs.~(\ref{fwti}), (\ref{gwti}).

\begin{figure}[t]
\centerline{\includegraphics[height=8.5cm]{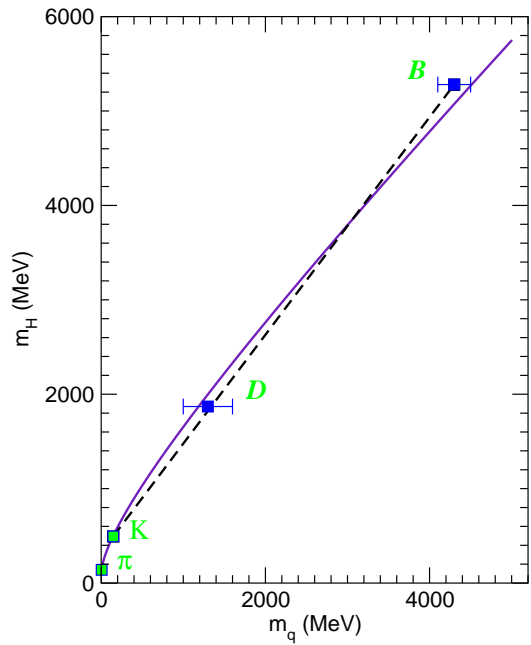}}\vspace*{-4ex}

\caption{\label{figmH} Solid line: pseudoscalar $u\bar q$ meson's mass as a
function of $m_q^{\zeta}$, $\zeta=19\,{\rm GeV}$, with a fixed value of
$m_u^{\zeta}$ that corresponds to $m_u^{1\,{\rm GeV}}=5.5\,$MeV,
Eq.~(\protect\ref{massuq}). The experimental data points are taken from
Ref.~\protect\cite{pdg00} as are the errors assigned to the associated
heavy-quark masses.  The dashed curve is a straight line drawn through the $K$,
$D$, $B$ masses.\vspace*{-4ex}}
\end{figure}

The unification of light- and heavy-meson masses via the mass formula in
Eq.~(\ref{massform}) has also been quantitatively explored using the model of
Refs.~\cite{mr97,pieterrho,pieterpion,pieterother}.  This is illustrated in
Fig.~\ref{figmH} wherein the calculated mass of a $u \bar q$ pseudoscalar meson
is plotted as a function of $m_q^\zeta$, with $m_u^\zeta$ fixed to a value
corresponding to that in Eq.~(\ref{massuq}).  The fitted curve is, in
MeV~\cite{pmqciv}:
\begin{equation}
\label{pietermH} m_H = 83 + 500 \sqrt{\cal X} + 310\,{\cal X},\,
{\cal X}= m_q^{\zeta}/\Lambda_{\rm QCD},
\end{equation}
with the renormalisation mass-scale $\zeta=19\,$GeV and $\Lambda_{\rm
QCD}=0.234\,$GeV.

In this figure the curvature appears slight but that is misleading: the
nonlinear term in Eq.~(\ref{pietermH}) accounts for almost all of $m_\pi$ (the
Gell-Mann--Oakes-Renner relation is nearly exact for the pion) and $80\,$\% of
$m_K$.  NB.\ The dashed line in Fig.~\ref{figmH} fits the $K$, $D$, $B$ subset
of the data exactly.  It is drawn to illustrate how easily one can be misled:
without careful calculation one might infer from this apparent agreement that
the large-$m_q$ limit of Eq.~(\ref{massform}) is already manifest at the
$s$-quark mass. However, in reality, the linear term only becomes dominant for
$m_q \gsim 1\,$GeV, providing $50\,$\% of $m_D$ and $67\,$\% of $m_B$.  The
model predicts $m_c^{1\,{\rm GeV}}=1.1\,$GeV and $m_b^{1\,{\rm GeV}}=4.2\,$GeV,
values that are typical of Poincar\'e covariant treatments of heavy-meson
systems~\cite{mishaheavy}.

A similar analysis of pseudoscalar mesons with equally massive constituents has
also been performed~\cite{pmqciv} and this predicts~\cite{cdrqciv}:
\begin{equation}
\frac{m_{H_{m=2 m_s}}}{m_{H_{m=m_s}}} = 2.2\,,
\end{equation}
in agreement with a result obtained in recent quenched lattice
simulations~\cite{michaels}.  The model calculation provides an understanding
of the lattice result.  It shows that the persistent dominance by the term
nonlinear in the current-quark mass owes itself to a large value of the
in-meson condensates for light-quark mesons; e.g., $-\langle \bar q
q\rangle^{s\bar s}_{1\,{\rm GeV}}= (0.32\,{\rm GeV})^3$~\cite{mr97}, and
thereby provides a confirmation of the large condensate version of chiral
perturbation theory.

\section{Pion's Valence Quark Distribution}
I have illustrated that the DSEs furnish a sound theoretical description of
pion properties, providing a model-independent explanation of its essential
dichotomous nature as both a Goldstone boson and a low-mass bound state of
massive constituents, and also that they supply an efficacious phenomenological
tool.  This makes them ideal for exploring the nature of the parton
distribution functions in the pion.  These functions, while they provide a
direct measure of the pion's quark-gluon substructure via perturbative QCD,
cannot be calculated using perturbative methods.

\begin{figure}[t]
\vspace*{1ex}

\centerline{\includegraphics[height=6.4cm]{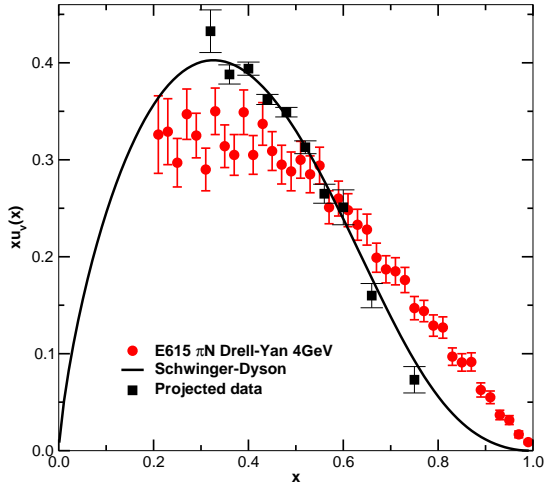}}\vspace*{-4ex}

\caption{\label{figuvx} DSE result for $x \,u^\pi_V(x)$~\protect\cite{uvxdse}
evolved to $\mu^2=16\,$GeV$^2$ using the first-order, nonsinglet
renormalisation group equation, for direct comparison with $\pi N$ Drell-Yan
data~\protect\cite{DY}, filled circles. The filled squares illustrate the
anticipated errors achievable in a proposed JLAB
experiment~\protect\cite{royJLAB}. \vspace*{-4ex}}
\end{figure}

The valence-quark distribution function, $u_V^\pi(x)$, is of particular
interest because its pointwise behaviour is affected by aspects of confinement
dynamics, e.g., by the mechanisms responsible for the finite extent and
essentially nonpointlike nature of the pion.  (NB.\ $u^{\pi^+} = \bar
d^{\pi^+}$ in the ${\cal G}$-parity symmetric limit of QCD.)  The low moments
of these functions can be inferred from numerical simulations of lattice-QCD
but such moments alone are insufficient to determine the functions: they are
only weakly sensitive to the behaviour of the distribution functions on the
valence-quark domain, $x\gsim 0.5$.

The DSE model used in illustrating the framework's ability to unify the small-
and large-$Q^2$ behaviour of the pion's form factors was used to calculate
$u_V^\pi(x)$~\cite{uvxdse}.  This calculation of the appropriate ``handbag
diagrams'' indicates that at a resolving scale $q_0=0.54\,$GeV$\,=1/(0.37
\,{\rm fm})$ valence quarks, with a mass of $0.3\,$GeV, carry $71\,$\% of the
pion's momentum, and yields the distribution function depicted in
Fig.~\ref{figuvx}.

Over the entire range of $q_0$ considered, the calculated distribution function
is precisely fitted by a MSR form
\begin{equation}
u_V^\pi(x) = A_u \, x^{\eta_1-1} \, (1-x)^{\eta_2} \, ( 1 - \epsilon_u \sqrt{x}
+ \gamma_u x),
\end{equation}
with exemplary, calculated parameter values:
\begin{equation}
\begin{array}{l|rrrrr}
q_0\,({\rm GeV}) & A_u & \eta_1 & \eta_2 & \epsilon_u & \gamma_u \\ \hline
0.54 & 11.24 & 1.43 & 1.90 & 2.44 & 2.54 \\
2.0 & 4.25 & 0.97 & 2.43 & 1.82 & 2.46 \\
4.05 & 3.56 & 0.89 & 2.61 & 1.62 & 2.30
\end{array}
\end{equation}
At $q_0=2.0\,$GeV two low moments of the distribution function
are~\cite{uvxdse}:
\begin{equation}
\begin{array}{l|lll}
  & \langle x^2 \rangle & \langle x^3 \rangle \\ \hline
{\rm Calc.} & 0.098 & 0.049 \\\hline
{\rm Exp.} & 0.10\pm 0.01 & 0.058\pm 0.004\\
{\rm Latt.}  & 0.11 \pm 0.3 & 0.048 \pm 0.020\\\hline
\end{array}
\end{equation}
with the experimental results from Ref.~\cite{DY} and the lattice results from
Ref.~\cite{lattice}.  NB.\ Given the evident disagreement between the DSE
calculation and the data, the agreement between the calculated moments
emphasises the insensitivity of low moments to the $x$-dependence of the
distribution function on the valence-quark domain.

A material feature of the DSE result is the value of $\eta_2 \simeq 2$ because
while perturbative QCD cannot be used to obtain the pointwise dependence of the
distribution functions it does give a prediction for the power-law dependence
at $x\simeq 1$.  That prediction is~\cite{uvxpQCD}:
\begin{equation}
\mbox{pQCD:}\;\;\; u_v^\pi(x) \stackrel{x\sim 1}{\propto} (1-x)^2\,,
\end{equation}
in agreement with the DSE result. However, as will have been anticipated from
Fig.~\ref{figuvx}, this prediction disagrees with the extant experimental
data~\cite{DY}, an analysis of which yields:
\begin{equation}
\mbox{Drell~Yan:}\;\;\;u_V^\pi(x) \stackrel{x\sim 1}{\propto} (1-x)\,.
\end{equation}
The disagreement is very disturbing because a verification of this experimental
result would present a profound threat to QCD, even challenging the assumed
vector-exchange nature of the force underlying the strong interaction.

\begin{figure}[t]
\centerline{\includegraphics[height=3.0cm]{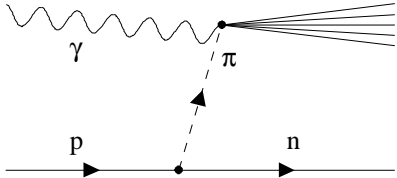}}\vspace*{-4ex}

\caption{\label{figsullivan} Deep inelastic scattering from the proton's
$\pi$-cloud could provide a means of measuring $u_V^\pi(x)$ at
JLAB~\protect\cite{royJLAB}.\vspace*{-4ex}}
\end{figure}

\begin{figure}[t]
\centerline{\includegraphics[height=4.5cm]{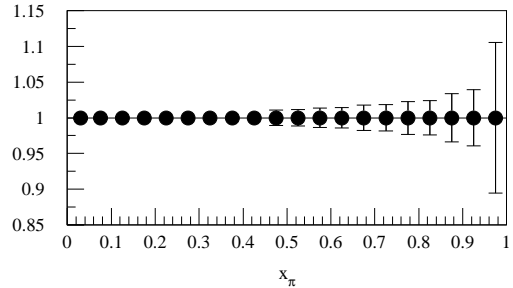}}\vspace*{-8ex}

\caption{\label{figerrors} Simulated errors for DIS events obtained in
collisions of a $5\,$GeV electron beam and a $25\,$GeV proton beam with a
luminosity of $10^{32}\,$cm$^{-2}\,$s$^{-1}$ and $10^6\,$s of running. (Figure
adapted from Ref.~\protect\cite{royEIC}.) \vspace*{-5ex}}
\end{figure}

The DSE study~\cite{uvxdse} has refocused attention on this disagreement, and
is the catalyst for a resurgence of interest in $u^\pi_V(x)$ and proposals for
its remeasurement.  One proposal that could use existing facilities would
employ the (Sullivan) process depicted in Fig.~\ref{figsullivan} at
JLAB~\cite{royJLAB}, with the anticipated accuracy illustrated in
Fig.~\ref{figuvx}.  This process could also be used efficaciously at a future
electron-proton collider to accurately probe $u_V^\pi(x)$ on the valence-quark
domain, as emphasised by Fig.~\ref{figerrors}~\cite{royEIC}.

\section{Epilogue}
Dynamical chiral symmetry breaking (DCSB) is a keystone of hadron physics. It
is the effect responsible for turning perturbative current-quark masses into
nonperturbative constituent-quark masses, and for ensuring that the pion is
light while the pion's electroweak decay constant nevertheless sets a large
mass-scale: $4 \pi f_\pi \sim 1\,$GeV.   It is thereby the foundation for the
successful application of chiral perturbation theory to low-energy hadronic
phenomena.

The QCD gap equation supplies a quark-level explanation of DCSB and, as one of
the tower of Dyson-Schwinger equations (DSEs), unifies that with a Poincar\'e
covariant understanding of the structure of QCD's bound states and their
interactions.  In this way one finds, for example, that DCSB is also the main
reason for the large $\pi-\rho$ and $\rho-a_1$ mass-splittings.

The fact that perturbation theory is recovered via a weak coupling expansion of
the DSEs provides a tight constraint on the ultraviolet behaviour of the
calculated Schwinger functions that describe hadronic interactions. Therefore
their use in predicting and correlating observables, which necessarily sees the
introduction of some model dependence, provides a means by which experimental
results can be used to probe the infrared (long-range) behaviour of the
quark-quark interaction; i.e., of exploring the mechanism of confinement.  In
this way the framework's adaptability to modelling is a material asset.

As reviewed in Refs.~\cite{revbasti,revreinhard}, modern applications of the
DSEs  have met with substantial success, even at nonzero temperature and
chemical potential, but challenges remain, of course. The gap equation makes
clear that the existence of DCSB signals a significant enhancement of the
strong coupling over the perturbative expectation on the infrared domain: $k^2
\lsim 1\,$GeV$^2$. However, the mechanism in QCD that supplies that enhancement
is yet to be conclusively identified, and is being explored using DSE and
lattice methods.

A direct bound state treatment of the scalar meson sector is also wanting.
However, at least one now understands why the lowest order nonperturbative
truncation of the kernels in the relevant integral equations (rainbow-ladder),
so successful for pseudoscalar and vector mesons, fails for the
scalars~\cite{cdrqcii}.  Improvements, systematic and/or imaginative, are being
explored.

This, after all, is a core issue in DSE studies and the primary point of
criticism: the system of integral equations must be truncated and how does one
judge, \textit{a priori}, the fidelity of a given procedure?  Addressing this
open question is a key focus of contemporary research but the question's
existence does not itself diminish the efficacy nor the value of modern DSE
phenomenology.


\end{document}